%% file: gdnp_2c.tex
\documentclass[twocolumn,english,aps,prc,superscriptaddress,byrevtex]{revtex4}
\usepackage[T1]{fontenc}
\usepackage[latin1]{inputenc}
\usepackage{amsmath}
\usepackage{graphicx}
\usepackage{amssymb}

\makeatletter

\usepackage{color}
\usepackage{graphicx}
\providecommand{\tabularnewline}{\\}
\usepackage{bm}

\input{my_macro}

\allowdisplaybreaks

\usepackage{babel}
\makeatother
\begin{document}

\title{Parity Nonconservation in the Photodisintegration of the Deuteron
at Low Energy}

\author{C.-P. Liu}

\email{C.P.Liu@KVI.nl}

\affiliation{TRIUMF, 4004 Wesbrook Mall, Vancouver, British Columbia, Canada V6T
2A3}

\affiliation{KVI, Zernikelaan 25, Groningen 9747 AA, The Netherlands}

\author{C. H. Hyun}

\email{hch@meson.skku.ac.kr}

\affiliation{School of Physics, Seoul National University, Seoul, 151-742, Korea}

\affiliation{Institute of Basic Science, Sungkyunkwan University, Suwon 440-746,
Korea}

\author{B. Desplanques}

\email{desplanq@lpsc.in2p3.fr}

\affiliation{Laboratoire de Physique Subatomique et de Cosmologie (UMR CNRS/IN2P3-UJF-INPG),\\
F-38026 Grenoble Cedex, France}

\date{\today{}}

\begin{abstract}
The parity-nonconserving asymmetry in the deuteron photodisintegration,
$\vec{\gamma}+d\rightarrow n+p$, is considered with the photon energy
ranged up to 10 MeV above the threshold. The aim is to improve upon
a schematic estimate assuming the absence of tensor as well as spin-orbit
forces in the nucleon-nucleon interaction. The major contributions
are due to the vector-meson exchanges, and the strong suppression
of the pion-exchange contribution is confirmed. A simple argument,
going beyond the observation of an algebraic cancellation, is presented.
Contributions of meson-exchange currents are also considered, but
found to be less significant. 
\end{abstract}
\maketitle

\section{Introduction \label{sec:intro}}

Some interest, both experimental and theoretical, has recently been
shown for the study of parity nonconservation in the deuteron photodisintegration
by polarized light. Historically, it was its inverse counterpart:
the net polarization in radiative thermal neutron capture by proton,
$n+p\rightarrow d+\gamma$, which attracted the first attention \cite{Danilov:1965}.
The experimental study was performed by the Leningrad group, taking
advantage of new techniques measuring an integrated current \cite{Lobashov:1972}.
The non-zero polarization obtained, $P_{\gamma}=-(1.3\pm0.45)\times10^{-6}$,
motivated many theoretical calculations in the frame of strong and
weak interaction models known in the 70's (see for instance Refs.
\cite{Lassey:1975,Desplanques:1975,Craver:1976am}). The theoretical
results were consistently within the range $P_{\gamma}=(2\sim5)\times10^{-8}$,
which is smaller than the measurement by a factor of 30 or more in
magnitude and, moreover, of opposite sign. The difficulty to understand
the measurement and, also perhaps, the novelty of the techniques,
which have been extensively used later on, led to a special reference
to this work as {}``Lobashov experiment''.

Later estimates with modern nucleon-nucleon ($NN$) potentials, both
parity-conserving (PC) and parity-nonconserving (PNC), give values
of $P_{\gamma}$ roughly within the same theoretical range as above.
On the experimental side, new results were reported in the early 80's
by the same Leningrad group, giving $P_{\gamma}\leq5\times10^{-7}$
\cite{Knyazkov:1983ke} and $P_{\gamma}=(1.8\pm1.8)\times10^{-7}$
\cite{Knyazkov:1984}. Practically, these results indicate an upper
limit of $P_{\gamma}$, which is not very constrictive. Since Leningrad
group's last report, the {}``Lobashov experiment'' has long been
forgotten by both experimentalists and theorists. Recent experiments
such as elastic $\vec{p}$-$p$ scattering (TRIUMF \cite{Berdoz:2001nu})
and polarized thermal neutron capture by proton (LANSCE \cite{Snow:2000az}),
which directly address the problem of PNC $NN$ interactions; and
quasi-elastic $\vec{e}$-$d$ scattering (MIT-Bates \cite{SAMPLE00b,Ito:2003mr}),
which indirectly involves these interactions, have however raised
a new interest for the study of PNC effects in few-body systems. In
what could be a golden age for these studies, the {}``Lobashov experiment''
is again evoked.

While it seems that there is not much prospect for performing the
{}``Lobashov experiment'' in a near future, the inverse process,
on the contrary, could be more promising. In this reaction, $\vec{\gamma}+d\rightarrow n+p$,
where a deuteron is disintegrated by absorbing a circularly polarized
photon, it is expected that, near threshold, the PNC asymmetry ($A_{\gamma}$)
is equal to the polarization in the {}``Lobashov experiment''. This
last one can thus be tested from a different approach.

The asymmetry $A_{\gamma}$ in the deuteron photodisintegration was
first calculated by Lee \cite{Lee:1978kh} up to the photon energy
$\omega_{\gamma}\simeq3.22$ MeV, which is 1 MeV above the threshold.
In this energy domain, where the dominant regular transition is $M1$,
the result was within the theoretical range of $P_{\gamma}$. Later
on, Oka extended Lee's work, up to $\omega_{\gamma}\simeq35$ MeV
\cite{Oka:1983sp}. Though the cross section still receives a contribution
from the $M1$ transition, the dominant contribution comes from the
$E1$ transitions. This offers a pattern of PNC effects different
from the one at very low photon energy. It was found that $A_{\gamma}$
shows a great enhancement at $\omega_{\gamma}\gtrsim5$ MeV, mainly
due to the PNC $\pi$-exchange contribution. If such an enhancement
were observed in the experiment, it would provide an important and
unambiguous determination of the weak $\pi NN$ coupling constant
$h_{\pi}^{1}$. However, a recent schematic calculation of $A_{\gamma}$
by Khriplovich and Korkin \cite{Khriplovich:2000mb}, partly suggested
by one of the present author, showed critical contradiction to Oka's
result, with a huge suppression of $A_{\gamma}$ at the energies $\omega_{\gamma}\gtrsim3$
MeV.

On the experimental side, a measurement of the asymmetry $A_{\gamma}$
in $\vec{\gamma}+d\rightarrow n+p$ was considered in the 80's by
E. D. Earle \textit{et al.} \cite{Earle:1981,Earle:1988fc} but no
sensitive result was reported. However, due to advances in experimental
techniques and instrumentation, the measurement of $A_{\gamma}$ becomes
more feasible nowadays and several groups at JLab \cite{jlab-lett00},
IASA (Athens), LEGS (BNL), TUNL, and SPring-8 show interest in such
a measurement. It is therefore important to understand and improve
previous estimates.

In this work, we carefully re-examine the $\vec{\gamma}+d\rightarrow n+p$
process with two main purposes:

\begin{enumerate}
\item Determine how the enhancement of the $h_{\pi}^{1}$ contribution in
Oka's results will change when the calculation is completed with missing
parity-admixed components in the final state, in particular in the
$^{3}P_{1}$ channel. The role of this last one was revealed by the
schematic estimate of Ref. \cite{Khriplovich:2000mb}. 
\item Determine the uncertainty of Khriplovich and Korkin's calculation
in which very simple wave functions are used. 
\end{enumerate}
It is straightforward to deal with the point 1. In Ref. \cite{Khriplovich:2000mb},
a nice and simple argument about the cancellation of the $h_{\pi}^{1}$
contribution from the final $^{3}P_{0}$, $^{3}P_{1}$, and $^{3}P_{2}$
states along with their parity-admixed partners was given. However,
the argument assumed the absence of tensor as well as spin-orbit forces,
which are important components of the $NN$ interaction. In order
to address these two points (missing components and simplicity of
the wave functions), we elaborate our calculation with the Argonne
$v_{18}$ $NN$ interaction model. We thus include the $^{1}S_{0}$,
$^{3}P_{0}$, $^{3}P_{1}$, $^{3}P_{2}$--$^{3}F_{2}$ channels, deuteron
$D$-state, and all their parity-admixed partners consistently. They
represent a minimal set of states that allows one to verify the results
of the schematic model as well as to include the effect of the tensor
and spin-orbit forces that manifest differently in these various channels.
We also include other channels, whose role is less important however.
As for the $E1$ operator, we employ the Siegert's theorem \cite{Siegert:1937yt},
which takes into account the contribution of some PC and PNC two-body
currents. The small photon energy considered here ($\omega_{\gamma}\leq12$
MeV) justifies this usage. Since there is no theorem similar to the
Siegert one for the $M1$ transition operator, two-body currents have
to be considered explicitly for both the PC and PNC parts. Adopting
Desplanques, Donoghue, Holstein (DDH) potential of the weak interaction
\cite{Desplanques:1980hn}, the asymmetry $A_{\gamma}$ will be expressed
in terms of the weak $\pi NN$, $\rho NN$ and $\omega NN$ coupling
constants, with corresponding coefficients indicating their relative
importance.

This paper is organized as follows. In Sect. \ref{sec:formalism},
we review the basic formalism underlying the calculation, which involves
both one- and two-body currents. In Sect. \ref{sec:results}, we show
the results and some discussions follow. A particular attention is
given to a comparison with earlier works and to new contributions
from PNC two-body currents. A simple argument explaining the suppression
of the pion-exchange contribution is also given. Conclusions are given
in Sect. \ref{sec:conclusion}. An appendix contains expressions of
$E1$ and $M1$ transition amplitudes due to the PNC two-body currents
considered in the present work.

\section{Formalism\label{sec:formalism}}

For a photodisintegration of an unpolarized target, the asymmetry
factor is defined as \[
A_{\gamma}\equiv\frac{\sigma_{+}-\sigma_{-}}{\sigma_{+}+\sigma_{-}}\,,\]
 where $\sigma_{+(-)}$ denotes the total cross section using right-
(left-) handed polarized light. By spherical multipole expansion,
it could be expressed as\begin{equation}
A_{\gamma}=\frac{{\displaystyle 2\,\mathrm{Re}\,\sum_{f,i,J}\,\left[F_{EJ}^{*}\,\tilde{F}_{MJ_{5}}+F_{MJ}^{*}\,\tilde{F}_{EJ_{5}}\right]}}{{\displaystyle \sum_{f,i,J}\,\left[F_{EJ}^{2}+F_{MJ}^{2}\right]}}\,.\label{eq:asym}\end{equation}
In this formula, the normal electromagnetic (EM) and PNC-induced EM
form factors, $F_{XJ}$ and $\tilde{F}_{XJ_{5}}$, with $X$ and $J$
denoting the type and multipolarity of the transition between a specific
initial ($i$) and final ($f$) states, are defined in the same way
as Refs. \cite{Musolf:1994tb,Liu:2002bq}. They depend on the momentum
transfer $q$, which equals to the photon energy $\omega_{\gamma}$
in this current case. The form factors $\tilde{F}_{XJ_{5}}$ (and
so does the asymmetry) vanish unless some PNC mechanism induces parity
admixtures of wave functions and axial-vector currents.

In this work, we consider the photon energy $\omega_{\gamma}=q$ up
to 10 MeV above the threshold. As the long wavelength limit, $\langle q\, r\rangle\ll1$,
is a good approximation, the inclusion of only dipole transitions,
\emph{i.e.} $E1$ and $M1$, is sufficient. This leads to 10 possible
exit channels connected to the deuteron state by angular momentum
considerations. Among them, $^{1}S_{0}$, via the $M1$ transition,
and $^{3}P_{0}$, $^{3}P_{1}$, $^{3}P_{2}$--$^{3}F_{2}$, via the
$E1$ transitions, dominate the cross section.

The transverse multipole operators assume a full knowledge of nuclear
currents. This requires, besides the one-body current $\bm j^{(1)}$
from individual nucleons, a complete set of two-body exchange currents
(ECs) $\bm j^{(2)}$ which is consistent with the nucleon-nucleon
($NN$) potential. These ECs are usually the sources of theoretical
uncertainties, because the $NN$ dynamics is still not fully understood.
While there is no alternative for the evaluation of $F_{MJ}$, the
Siegert theorem \cite{Siegert:1937yt} does allow one to transform
the evaluation of $F_{EJ}$ into the one of charge multipole $F_{CJ}$.
The fact that the PC $NN$ interaction does not give rise to exchange
charges at $O(1)$ removes most of the uncertainties related to exchange
effects: knowledge of the one-body charge $\rho^{(1)}$ is sufficient
for a calculation good to the order of $1/\mN$.

In the framework of impulse approximation and using the Siegert theorem,
one gets, for the deuteron photodisintegration ($E_{f}-E_{i}=\omega_{\gamma}=q$
and $J_{i}=1$),\begin{widetext}

\begin{eqnarray}
F_{E1}^{(S)}(q)_{f,i} & = & \frac{E_{i}-E_{f}}{q}\,\sqrt{\frac{2}{2J_{i}+1}}\,\bra{J_{f}}|\int\, d^{3}x\,[j_{1}(q\, x)\, Y_{1}(\Omega_{x})]\,\rho^{(1)}(\bm x)|\ket{J_{i}}\nonumber \\
 &  & +\frac{1}{q}\,\frac{1}{\sqrt{2J_{i}+1}}\,\bra{J_{f}}|\int\, d^{3}x\,\bm\nabla\times[j_{1}(q\, x)\,\bm Y_{111}(\Omega_{x})]\cdot\bm j_{spin}^{(1)}(\bm x)|\ket{J_{i}}\nonumber \\
 & \simeq & -\frac{q}{3\sqrt{2\,\pi}}\,\bra{J_{f}}|\sum_{i}\,\hat{e}_{i}\,\bm x_{i}|\ket{J_{i}}\equiv-\frac{q}{2\,\sqrt{6\,\pi}}\,\langle E1^{(1)}\rangle\,,\label{eq:Siegert E1}\\
F_{M1}^{(1)}(q)_{f,i} & = & i\,\frac{1}{\sqrt{2J_{i}+1}}\,\bra{J_{f}}|\int\, d^{3}x\,[j_{1}(q\, x)\,\bm Y_{111}(\Omega_{x})]\cdot\bm j^{(1)}(\bm x)|\ket{J_{i}}\nonumber \\
 & \simeq & -\frac{q}{3\sqrt{2\,\pi}}\,\bra{J_{f}}|\sum_{i}\,\frac{1}{2\mN}\,[\hat{e}_{i}\,\bm x_{i}\times\bm p_{i}+\hat{\mu}_{i}\,\bm\sigma_{i}]|\ket{J_{i}}\,\equiv-\frac{q}{2\,\sqrt{6\,\pi}}\,\langle M1^{(1)}\rangle\,,\end{eqnarray}
\end{widetext}where $\hat{e}_{i}=e\,(1+\tau_{i}^{z})/2$ and $\hat{\mu}_{i}=e\,(\muS+\muV\tau_{i}^{z})/2$
with $\muS=0.88$ and $\muV=4.70$; $Y$ and $\bm Y$ are the spherical
and vector spherical harmonics. In these expressions, the approximated
results are obtained by replacing the spherical Bessel function $j_{1}(q\, x)$
with its asymptotic form as $q\rightarrow0$, \emph{i.e.} $q\, x/3$,
at the long wavelength limit and keeping terms linear in $q$ (the
lowest order); they could be related to the forms of $\langle E1^{(1)}\rangle$
and $\langle M1^{(1)}\rangle$ often adopted in the literature. In
our numerical calculation, the identity relations are employed instead.
Note that the one-body spin current is conserved by itself and not
constrained by current conservation. In Eq. (\ref{eq:Siegert E1}),
this one-body spin current (2nd line) is of higher order in $q$ compared
with the Siegert term (1st line), however, it is kept for completeness.
As for the PNC-induced form factors $\tilde{F}_{E1_{5}}^{(S)}$ and
$\tilde{F}_{M1_{5}}^{(1)}$, one only has to replace either the initial
or final state by its opposite-parity admixture, $\widetilde{{\bra{J_{f}}}}$
or $\widetilde{{\ket{J_{i}}}}$, and add a factor {}``$i$'' for
$E1$ or {}``$-i$'' for $M1$ matrix elements (in relation with
our conventions).

The non-vanishing matrix elements for the five dominant exit channels
are thus\begin{widetext}

\begin{enumerate}
\item $^{1}S_{0}$:\begin{eqnarray}
\langle M1^{(1)}\rangle & = & -\frac{\muV}{\mN}\,\int\, dr\, U^{*}(^{1}S_{0})\, U_{d}(^{3}S_{1})\,,\label{eq:comp-i}\\
\langle E1_{5}^{(1)}\rangle & = & \frac{i}{3}\,\int\, r\, dr\,\tilde{U}^{*}(^{3}P_{0})\,\left[U_{d}(^{3}S_{1})-\sqrt{2}\, U_{d}(^{3}D_{1})\right]\nonumber \\
 &  & -\frac{i}{\sqrt{3}}\,\int\, r\, dr\, U^{*}(^{1}S_{0})\,\tilde{U}_{d}(^{1}P_{1})\,.\end{eqnarray}

\item $^{3}P_{0}$:\begin{eqnarray}
\langle E1^{(1)}\rangle & = & \frac{1}{3}\,\int\, r\, dr\, U^{*}(^{3}P_{0})\,\left[U_{d}(^{3}S_{1})-\sqrt{2}\, U_{d}(^{3}D_{1})\right]\,,\\
\langle M1_{5}^{(1)}\rangle & = & i\,\frac{\muV}{\mN}\,\int\, dr\,\left[\tilde{U}^{*}(^{1}S_{0})\, U_{d}(^{3}S_{1})-\frac{1}{\sqrt{3}}\, U^{*}(^{3}P_{0})\,\tilde{U}_{d}(^{1}P_{1})\right]\nonumber \\
 &  & -\, i\,\sqrt{\frac{2}{3}}\,\frac{\muS-1/2}{\mN}\,\int\, dr\, U^{*}(^{3}P_{0})\,\tilde{U}_{d}(^{3}P_{1})\,.\label{eq:comp-ia}\end{eqnarray}

\item $^{3}P_{1}$:\begin{eqnarray}
\langle E1^{(1)}\rangle & = & -\frac{1}{\sqrt{3}}\,\int\, r\, dr\, U^{*}(^{3}P_{1})\,\left[U_{d}(^{3}S_{1})+\frac{1}{\sqrt{2}}\, U_{d}(^{3}D_{1})\right]\,,\\
\langle M1_{5}^{(1)}\rangle & = & -\, i\,\frac{\muV}{\mN}\,\int\, dr\, U^{*}(^{3}P_{1})\,\tilde{U}_{d}(^{1}P_{1})-i\,\frac{\muS+1/2}{\sqrt{2}\,\mN}\,\int\, dr\, U^{*}(^{3}P_{1})\,\tilde{U}_{d}(^{3}P_{1})\nonumber \\
 &  & -\, i\,\frac{\sqrt{2}\,\muS}{\mN}\,\int\, dr\,\tilde{U}^{*}(^{3}S_{1})\, U_{d}(^{3}S_{1})+i\,\frac{\muS-3/2}{\sqrt{2}\,\mN}\,\int\, dr\,\tilde{U}^{*}(^{3}D_{1})\, U_{d}(^{3}D_{1})\,.\label{eq:comp-ib}\end{eqnarray}

\item $^{3}P_{2}$--$^{3}F_{2}$:\begin{eqnarray}
\langle E1^{(1)}\rangle & = & \frac{\sqrt{5}}{3}\,\int\, r\, dr\,\bigg\{ U^{*}(^{3}P_{2})\,\left[U_{d}(^{3}S_{1})-\frac{1}{5\,\sqrt{2}}\, U_{d}(^{3}D_{1})\right]\,\nonumber \\
 &  & +\frac{3\,\sqrt{3}}{5}\, U^{*}(^{3}F_{2})\, U_{d}(^{3}D_{1})\bigg\}\,,\\
\langle M1_{5}^{(1)}\rangle & = & -\, i\,\sqrt{\frac{5}{3}}\,\frac{\muV}{\mN}\,\int\, dr\,\left[U^{*}(^{3}P_{2})\,\tilde{U}_{d}(^{1}P_{1})-\sqrt{\frac{3}{5}}\,\tilde{U}^{*}(^{1}D_{2})\, U_{d}(^{3}D_{1})\right]\nonumber \\
 &  & +i\,\sqrt{\frac{5}{6}}\,\frac{\muS-1/2}{\mN}\,\int\, dr\,\left[U^{*}(^{3}P_{2})\,\tilde{U}_{d}(^{3}P_{1})+\frac{3}{\sqrt{5}}\,\tilde{U}^{*}(^{3}D_{2})\, U_{d}(^{3}D_{1})\right]\,.\label{eq:comp-f}\end{eqnarray}

\end{enumerate}
\end{widetext}Results for the remaining five less important channels
($^{3}S_{1}$--$^{3}D_{1}$, $^{1}P_{1}$, $^{1}D_{2}$, $^{3}D_{2}$)
will be included in numerical works. The $r$-weighted radial wave
functions for scattering and deuteron states, $U$ and $U_{d}$, along
with their parity admixtures, $\tilde{U}$ and $\tilde{U}_{d}$, are
obtained by solving the Schr\"{o}dinger equations. Details could
be found in Ref. \cite{Liu:2002bq}.

By taking the square of normal EM form factors (PC response function)
or the product of normal and PNC-induced ones (PNC response function),
we can directly compare Eqs. (5a--5h) in Ref. \cite{Oka:1983sp}.
After removing factors due to wave-function normalizations, the differences
are:

\begin{enumerate}
\item The parity admixture of the scattering $^{3}P_{1}$ state is included
in our work: The admixtures $\tilde{U}(^{3}S_{1})$ and $\tilde{U}(^{3}D_{1})$
are solved from the inhomogeneous differential equations with the
source term modulated by $U(^{3}P_{1})$. They are not orthogonal
to the deuteron state and thus should not be ignored. Actually, they
are required to ensure the orthogonality of the deuteron and the $^{3}P_{1}$
scattering states once these ones are allowed to contain a parity-nonconserving
component. 
\item The terms involving the scalar magnetic moment are different: Looking
for instance at the $M1$ matrix element between $U(^{3}P_{1})$ and
$\tilde{U}_{d}(^{3}P_{1})$, the effective $M1$ operator is proportional
to $\muS\,\bm S+\bm L/2$. By the projection theorem, $\langle\bm S\rangle=\langle\bm L\rangle$,
the overall factor should be $\muS+1/2$, not $\muS+1$ as in Ref.
\cite{Oka:1983sp}.%
\footnote{We also note that unlike our notation, $\mu_{\ssst{S,V}}$ is used
to denote the anomalous magnetic moments in Ref. \cite{Oka:1983sp}.%
} It looks as if this work ignored the $1/2$ factor in front of the
$\bm L$ operator. 
\end{enumerate}
Both points involve the spin-conserving PNC interaction, which is
dominated by the pion exchange. Therefore, how these differences change
the sensitivity of $A_{\gamma}$ with respect to $h_{\pi}^{1}$ will
be elaborated in next section.

Now we discuss, in two steps, extra contributions due to ECs when
one tries to go beyond the impulse approximation together with the
Siegert-theorem framework. 

First, when PC ECs are included, their contribution to $M1$ matrix
elements, $F_{M1}^{(2)}$, definitely needs to be calculated. On the
other hand, as PC exchange charges are higher-order in the nonrelativistic
limit, $F_{E1}^{(S)}$ is supposed to take care of most two-body effects,
and the remaining contribution $\Delta F_{E1}^{(2)}$ can be safely
ignored. This argument also applies for the PNC-induced form factors
involving the PC ECs: one needs to consider $\tilde{F}_{M1_{5}}^{(2)}$
but can leave out $\Delta\tilde{F}_{E1_{5}}^{(2)}$. 

Second, the inclusion of PNC ECs, to the first order in weak interaction,
only affects the PNC-induced form factors. The contribution $\tilde{F}_{M1_{5}}^{(2')}$
is calculated by using the $M1$ operator constructed from the PNC
ECs and unperturbed wave functions (so we use a prime to remind the
difference from parity-admixture contributions). One special feature
of PNC ECs is that they do have exchange charges of $O(1)$ \cite{Liu:2003au}.
Therefore, one should include them in $\tilde{F}_{E1_{5}}^{(S')}$.

As a last remark, we note one advantage of nuclear PNC experiments
in processes like photodisintegration or radiative capture. The real
photon is {}``blind'' to the nucleon anapole moment, which could
contribute otherwise to PNC observables in virtual photon processes.
Because this P-odd T-even nucleon moment is still poorly constrained
both theoretically and experimentally, the interpretation of real-photon
processes, like the one considered here, is thus comparatively easier.

\section{Results and Discussions \label{sec:results}}

For practical purposes, we use the Argonne $v_{18}$ \cite{Wiringa:1995wb}
(A$v_{18}$) and DDH \cite{Desplanques:1980hn} potentials as the
PC and PNC $NN$ interactions, respectively. In comparison with earlier
works in the 70's or the 80's, a strong interaction model like A$v_{18}$
offers the advantage that the singlet-scattering length is correctly
reproduced, due to its charge dependence. Correcting results with
this respect is therefore unnecessary.

\begin{figure}
\includegraphics[%
  scale=0.3,
  angle=270]{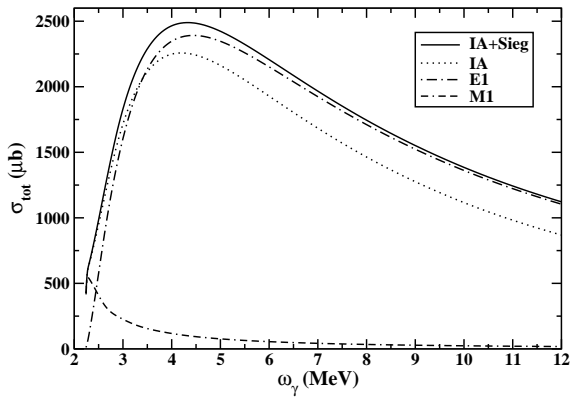}

\caption{The total cross section as a function of the photon energy. The main
result is the curve labeled as {}``IA+Sieg'', and the curves {}``E1''
and {}``M1'' showing contributions from corresponding transitions.
The curve {}``IA'' is the result of a pure impulse approximation
calculation, where no two-body contribution is included. \label{cap:cross section}}
\end{figure}

The total cross section is plotted in Fig. \ref{cap:cross section}
as a function of the photon energy and labeled as {}``IA+Sieg''.
Its separate contributions from $E1$ and $M1$ transitions are also
shown on the same plot (labeled accordingly). The $M1$ transition
only dominates near the threshold region; as the photon energy reaches
about 1 MeV above the threshold, the $E1$ transition overwhelms.
Away from the threshold, the calculated results agree well with both
experiment and existing potential-model calculations up to 10 MeV
\cite{Arenhovel:1991}. Such a good agreement shows the usefulness
of the Siegert theorem, by which most of the two-body effects are
included. Compared with the curve labeled by {}``IA'', the result
of impulse approximation, one sees the increasing importance of these
two-body contributions as $\omega_{\gamma}$ gets larger. On the contrary,
because $M1$ matrix elements are purely one-body, we expect our near-threshold
results smaller than experiment by about $10\%$ \cite{Arenhovel:1991}.
This discrepancy, originally found in the radiative capture of thermal
neutron by proton (the inverse of deuteron photodisintegration), requires
various physics such as exchange currents and isobar configurations,
to be fully explained. Here, we qualitatively estimate a $5\%$ error
for the calculation of $F_{M1}$ near threshold.

\begin{figure}
\includegraphics[%
  scale=0.3,
  angle=270]{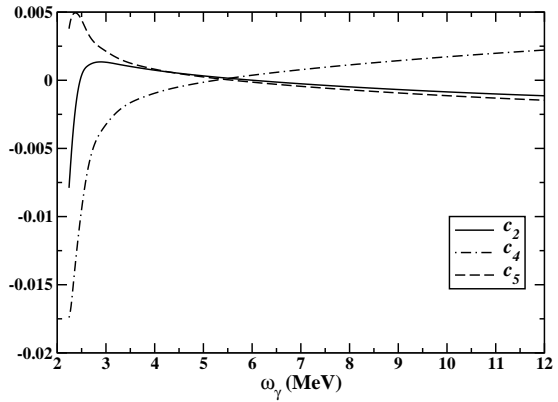}

\includegraphics[%
  scale=0.3,
  angle=270]{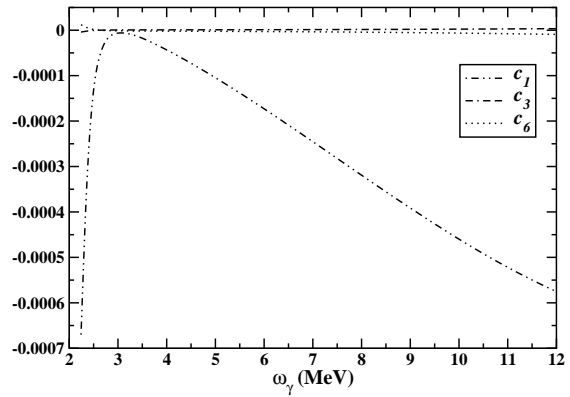}

\caption{The energy-dependences of {}``large'' coefficients $c_{2}$, $c_{4}$,
and $c_{5}$ (top panel) and {}``small'' coefficients $c_{1}$,
$c_{3}$ , and $c_{6}$ (bottom panel) in the asymmetry parametrization,
Eq. (\ref{eq:A exp}). \label{cap:coeffs}}
\end{figure}

When calculating the PNC-induced matrix elements with the DDH potential,
we use the strong meson-nucleon coupling constants: $g_{\pi\ssst{NN}}=13.45$,
$g_{\rho\ssst{NN}}=2.79$, and $g_{\omega\ssst{NN}}=8.37$, and meson
masses (in units of MeV): $m_{\pi}=139.57$, $m_{\rho}=770.00$, and
$m_{\omega}=781.94$. The resulting asymmetry is then expressed in
terms of six PNC meson-nucleon coupling constants $h$'s as \begin{equation}
A_{\gamma}=c_{1}\, h_{\pi}^{1}+c_{2}\, h_{\rho}^{0}+c_{3}\, h_{\rho}^{1}+c_{4}\, h_{\rho}^{2}+c_{5}\, h_{\omega}^{0}+c_{6}\, h_{\omega}^{1}\,,\label{eq:A exp}\end{equation}
 where the six energy-dependent coefficients $c_{1...6}$ show the
sensitivity to each corresponding coupling. It turns out that, for
the energy range considered here, $c_{2},\, c_{4},\, c_{5}\gg c_{1}\gg c_{3},\, c_{6}$.
This implies the asymmetry has a larger sensitivity to the isoscalar
and isotensor couplings than to the isovector ones. The detailed energy
dependences of these {}``large'' and {}``small'' coefficients
are shown in Fig. \ref{cap:coeffs}.

In principle, these results are independent. In practice however,
they can be shown to depend on three quantities, reflecting the dominant
role of the various $S\leftrightarrow P$ neutron-proton transition
amplitudes at low energy. These amplitudes have some energy dependence
which is essentially determined by the best known long-range properties
of strong interaction models. They can therefore be parametrized by
their values at zero energy \cite{Danilov:1965,Missimer:1976wb,Desplanques:1978mt},
including at the deuteron pole. To a large extent, they can be used
independently of the underlying strong interaction model, quite in
the spirit of effective-field theories that they anticipated \cite{Holstein}.
In the case of the A$v_{18}$ model employed here, they are given
by: \begin{eqnarray}
\mN\,\lambda_{t} & = & -0.043\, h_{\rho}^{0}-0.022\, h_{\omega}^{0}\,,\nonumber \\
\mN\,\lambda_{s} & = & -0.125\, h_{\rho}^{0}-0.109\, h_{\omega}^{0}+0.102\, h_{\rho}^{2}\,,\nonumber \\
\mN\, C & = & 1.023\, h_{\pi}^{1}+0.007\, h_{\rho}^{1}-0.021\, h_{\omega}^{1}\,.\end{eqnarray}
 The largest corrections to the above approach occur for the PNC pion-exchange
interaction which, due to its long range, produces some extra energy
dependence and sizable $P\leftrightarrow D$ transition amplitudes.
They can show up when the contribution of the $S\leftrightarrow P$
transition amplitude is suppressed, like in this work.

For $\omega_{\gamma}=2.235$ MeV, which is very close to the disintegration
threshold, we get the asymmetry \begin{eqnarray}
A_{\gamma}^{(th)} & \approx & \left[-8.44\, h_{\rho}^{0}-17.65\, h_{\rho}^{2}+3.63\, h_{\omega}^{0}\right.\nonumber \\
 &  & \left.\hspace{2cm}+O(c_{1},c_{3},c_{6})\right]\times10^{-3}\,.\label{eq:A-threshold}\end{eqnarray}
 Using the DDH {}``best'' values as an estimate, we got $A_{\gamma}^{(th)}\approx2.53\times10^{-8}$.
By detailed balancing, one expects that $A_{\gamma}^{(th)}$ equals
the circular polarization $P_{\gamma}^{(th)}$ observed in the radiative
thermal neutron capture by proton, given the same kinematics. Though
our result does not exactly correspond to the same kinematics as the
inverse process usually considered (the kinetic energy of thermal
neutrons $\sim$0.025 eV), it agrees both in sign and order of magnitude
with existing calculations of $P_{\gamma}^{(th)}$ \cite{Lassey:1975,Desplanques:1975,Craver:1976am}.
We also performed a similar calculation for the latter case with A$v_{18}$,
and the result is \begin{eqnarray}
P_{\gamma}^{(th)} & \approx & \left[-8.75\, h_{\rho}^{0}-17.47\, h_{\rho}^{2}+3.39\, h_{\omega}^{0}\right.\nonumber \\
 &  & \left.\hspace{2cm}+O(c_{1},c_{3},c_{6})\right]\times10^{-3}\,.\end{eqnarray}
 This is very close to the result of $A_{\gamma}$ quoted above.

It is noticed that our expression of $A_{\gamma}^{(th)}$ at very
low energy, and therefore that one for $P_{\gamma}^{(th)}$, contains
a contribution from the one-pion exchange (see the low-energy part
of the $c_{1}$ coefficient given in Fig. \ref{cap:coeffs}). This
feature, which apparently contradicts the statement often made in
the past that this contribution is absent in $P_{\gamma}^{(th)}$,
is due to the incorporation in our work of the spin term in Eq. (\ref{eq:Siegert E1}),
which represents a higher order term in $q$. This correction also
explains the difference in the behavior of the $c_{1}$ coefficient
with the Oka's result \cite{Oka:1983sp}.

We note that, because the $M1$ transition dominates at the threshold
and we only use the impulse approximation for its matrix element,
there should be approximately a $-5\%$ correction to $A_{\gamma}^{(th)}$
(also $P_{\gamma}^{(th)}$) when two-body effects are included in
$F_{M1}$. On the other hand, as $\tilde{F}_{E1_{5}}$ is calculated
using the Siegert theorem, it should be reliable up to the correction
of $\tilde{F}_{E1_{5}}^{(S')}$ from the PNC exchange charge at $O(1)$.

\begin{figure}
\includegraphics[%
  scale=0.3,
  angle=270]{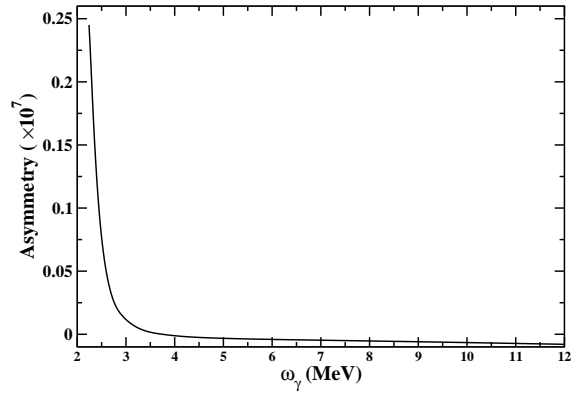}

\caption{The asymmetry by using the DDH best values. \label{cap:Ag DDH}}
\end{figure}

When the photon energy gets larger, one can see immediately that the
asymmetry gets smaller. A prediction using the DDH best values is
shown in Fig. \ref{cap:Ag DDH}. In this figure, as soon as the photon
energy reaches 1 MeV above the threshold, the asymmetry drops by one
order of magnitude. Moreover, the sign changes around $\omega_{\gamma}=4$
MeV. This implies that a higher sensitivity ($\sim10^{-9}$) is needed
for any experiment targeting at the kinematic range away from the
threshold. Our calculation is consistent with the work by Khriplovich
and Korkin \cite{Khriplovich:2000mb}, but is widely different from
the one by Oka \cite{Oka:1983sp}. In the following, we make a closer
comparison with these works and, then, present results for the contribution
of various PNC two-body currents considered for the first time.

\subsection{Comparison with Oka's work \label{sub:CompOka}}

\noindent The major difference comes from the pion sector. In Ref.
\cite{Khriplovich:2000mb}, where the scattering wave functions are
obtained from the zero-range approximation and the deuteron is purely
a $^{3}S_{1}$ state, a simple angular momentum consideration leads
to a null contribution from pions. Our result shows that the more
complex nuclear dynamics has only small corrections, so the asymmetry
is not sensitive to $h_{\pi}^{1}$. However, it is not the case at
all in Ref. \cite{Oka:1983sp}: the pion exchange dominates the asymmetry
with the coefficient $c_{1}$ being one or two orders of magnitude
larger than our result.

This discrepancy could be illustrated by considering a case where
$\omega_{\gamma}$ is 10 MeV above the threshold. In the central column
of Table \ref{cap:compare Oka's}, we list the pertinent PNC responses
due to the pion exchange among the 5 dominant exit channels. In the
right column, we simulate what the outcome will be if the analytical
results of Eqs. (5a--5g) in Ref. \cite{Oka:1983sp} are used, \emph{i.e.}
with different factors involving $\mu_{\ssst{S}}$ and no parity admixture
of $^{3}P_{1}$ state as mentioned in Sec. \ref{sec:formalism}. Comparing
the totals from both columns, one immediately observes the simulated
result is bigger by an order of magnitude. More inspection shows that,
while the changes of the $\mu_{\ssst{S}}$ factors do alter each response
somewhat, the major difference depends on whether the big cancellation
from the $^{3}P_{1}$ admixture is included or not. By adding contributions
from other sub-leading channels, the total will be further downed
by a factor of 2.5. Thus the overall difference is about a factor
of 30. %
\begin{table}
\begin{ruledtabular}

\begin{tabular}{ccc} Transitions& Eqs. (\ref{eq:comp-i}--\ref{eq:comp-f})& Eqs. (5a--5h) in Ref. \cite{Oka:1983sp}\tabularnewline \hline \hline $^{3}P_{0}\leftrightarrow\tilde{\mathcal{D}}$& 0.449& -0.142\tabularnewline $^{3}P_{1}\leftrightarrow\tilde{\mathcal{D}}$& -3.217& -4.383\tabularnewline $\widetilde{^{3}P_{1}}\leftrightarrow\mathcal{D}$& 3.942& not considered\tabularnewline $^{3}P_{2}\leftrightarrow\tilde{\mathcal{D}}$& -1.231& 0.389\tabularnewline $\widetilde{^{3}P_{2}}\leftrightarrow\mathcal{D}$& -0.142& 0.045\tabularnewline $^{3}F_{2}\leftrightarrow\tilde{\mathcal{D}}$& -0.151& 0.048\tabularnewline $\widetilde{^{3}F_{2}}\leftrightarrow\mathcal{D}$& -0.019& 0.006\tabularnewline \hline Total& -0.371& -4.037\tabularnewline \end{tabular}

\end{ruledtabular}

\caption{The dominant PNC responses due to the pion exchange for $\omega_{\gamma}$
10 MeV above the threshold (in units of $10^{-5}\times h_{\pi}^{1}$).
The central column is calculated by Eqs. (\ref{eq:comp-i}--\ref{eq:comp-f}),
while the right column by Eqs. (5a--5h) in Ref. \cite{Oka:1983sp}.
The symbol $\mathcal{D}$ denotes the deuteron state. \label{cap:compare Oka's}}
\end{table}

\subsection{\textit{\emph{Comparison with Khriplovich and Korkin's work \label{sub:CompKK}}}}

\noindent The vanishing of the $\pi$-exchange contribution in Khriplovich
and Korkin's work \cite{Khriplovich:2000mb} supposes that the $E1$
transitions from the deuteron state to the different scattering states,
$^{3}P_{0}$, $^{3}P_{1}$ and $^{3}P_{2}$, are the same, which implies
that one neglects both the tensor and spin-orbit components of the
strong interaction. As these parts of the force have large effects
in some cases, it is important to determine how the above vanishing
is affected when a more realistic description of the interaction is
used.

We first notice that the isoscalar magnetic operator, $\muS\,\bm S+\bm L/2$,
can be written as $\muS\,\bm J+(1/2-\muS)\bm L$. As the operator
$\bm J$ conserves the total angular momentum, it follows that the
$E1$ transitions from the deuteron state to the $^{3}P_{0}$ and
$^{3}P_{2}$ states will be proportional to $\muS-1/2$, in agreement
with Eqs. (\ref{eq:comp-ia}) and (\ref{eq:comp-f}). A similar result
holds for the $^{3}P_{1}$ state. For this transition, one has to
take into account that the $\bm J$ operator connects states that
are orthogonal to each other, including the case where they contain
some parity admixture. This unusual but interesting result was originally
suggested by a similar result obtained by Khriplovich and Korkin for
the $^{1}S_{0}$ and $^{3}P_{0}$ states \cite{Khriplovich:2000mb}.
They used it later on for the $\pi$-exchange contribution on the
suggestion of one the present authors. Taking this property into account,
one can check that the different $\muS$-dependent terms in Eq. (\ref{eq:comp-ib})
combine so that the quantity, $\muS-1/2$, can be factored out. This
explains the cancellation of the two largest contributions in Table
\ref{cap:compare Oka's}, $3.942$ and $-3.217$, approximately proportional
to $2\,\muS=1.76$ and $-(\muS+1/2)=-1.38$.

Further cancellation is obtained when one considers the sum of the
$\pi$-exchange contributions to the asymmetry $A_{\gamma}$ corresponding
to the different $P$ states. Taking into account the remark made
in the previous paragraph, it can be checked that contributions from
Eqs. (\ref{eq:comp-ia}), (\ref{eq:comp-ib}) and (\ref{eq:comp-f})
are proportional to 2, 3 and $-5$ and 4, $-3$, and $-1$ for the
$^{3}S_{1}$ and $^{3}D_{1}$ deuteron components respectively (assuming
that the $^{3}P$ wave functions are the same). As can be seen in
Table \ref{cap:compare Oka's}, the dominant contributions, 0.449,
0.725 $(=3.942-3.217)$ and $-1.231$ are not far from the relative
ratios 2, 3 and $-5$, expected for the $^{3}S_{1}$ deuteron component.
Possible departures can be ascribed in first place to the $^{3}D_{1}$
deuteron component.

The above cancellation calls for an explanation deeper than the one
consisting in the verification that the algebraic sum of different
contributions cancels. An argument could be the following. In the
conditions where the cancellation takes place (same interaction in
the $^{3}P$ states in particular), a closure approximation involving
spin and angular orbital momentum degrees of freedom can be used to
simplify the writing of the PNC part of the response function that
appears at the numerator of Eq. (\ref{eq:asym}). Keeping only the
factors of interest here, the interference term of $E1$ and $M1$
matrix elements can be successively transformed as follows \begin{eqnarray}
\delta R & \propto & \sum_{M}\bra{J_{i}}\,\hat{r}^{i}\,(\muS-\tfrac12)\, L^{j}\,\left(\delta^{ij}-\hat{q}^{i}\,\hat{q}^{j}\right)\widetilde{{\ket{J_{i}}}}\nonumber \\
 & \propto & \sum_{M}\Big[\bra{^{3}S_{1}}U_{d}(^{3}S_{1})+\frac{U_{d}(^{3}D_{1})}{\sqrt{2}}\,\left(3\,(\bm S\!\cdot\!\hat{r})^{2}-S^{2}\right)\,\Big]\nonumber \\
 &  & \hspace{1cm}\times\:\hat{r}^{i}\, L^{j}\,\left(\delta^{ij}-\hat{q}^{i}\,\hat{q}^{j}\right)\Big[\bm S\cdot\hat{r}\,\ket{^{3}S_{1}}\Big]\nonumber \\
 & \propto & {\textrm{Tr}}\left(\Big[U_{d}(^{3}S_{1})+\frac{U_{d}(^{3}D_{1})}{\sqrt{2}}\,\left(3\,(\bm S\!\cdot\!\hat{r})^{2}-S^{2}\right)\,\Big]\,\bm S\!\cdot\!\hat{r}\right)\nonumber \\
 & = & 0\,.\label{eq:cancellation}\end{eqnarray}
 The first line stems from retaining the isoscalar part of the magnetic
operator proportional to $(\muS-1/2)\bm L$ (it is reminded that the
$\bm J$ part does not contribute). The next line is obtained by expressing
the PC and PNC parts of the deuteron wave function as some operator
acting on a pure $|^{3}S_{1}\rangle$ state. Once this transformation
is made, it is possible to replace the summation over the deuteron
angular momentum components, $M$, by the spin ones, $m_{s}$, which
is accounted for at the third line. The last line then follows from
the fact that the trace of the spin operator, $\bm S$, possibly combined
with a $\Delta S=2$ one, vanishes. A result similar to the above
one can be obtained for some contributions involving MECs. It is however
noticed that some corrections involving the spin-orbit force, or spin-dependent
terms in the E1 transition operator, which both contain an extra $\bm S$
factor in the above equation, could lead to a non-zero trace and therefore
to a relatively large correction. Of course, the above cancellation
relies on the fact that no polarization of the initial or final state
is considered. Had we looked at an observable involving such a polarization,
like the asymmetry in the capture of polarized thermal neutrons by
protons, the result will be quite different. As is well known, this
observable is dominated by the $\pi$-exchange contribution \cite{Danilov:1965}.

\subsection{\textit{\emph{Contributions of PNC ECs \label{sub:PNC-ECs}}}}

In Section \ref{sec:formalism}, the contributions of PNC ECs were
summarized in two additional PNC-induced form factors, $\tilde{F}_{E1_{5}}^{(S')}$
and $\tilde{F}_{M1_{5}}^{(2')}$. Now, we estimate these contributions
by considering only the dominant channels $^{1}S_{0}$, $^{3}P_{0}$,
$^{3}P_{1}$ and $^{3}P_{2}$--$^{3}F_{2}$. As $E1_{5}$ connects
states of same parity, only $^{1}S_{0}$ is allowed; therefore, $\tilde{F}_{E1_{5}}^{(S')}$
plays a more important role for $A_{\gamma}$ near the threshold.
On the other hand, $M1_{5}$ connects states of opposite parity, which
requires the other four channels, so $\tilde{F}_{M1_{5}}^{(2')}$
has more impact on $A_{\gamma}$ at higher energies. The full set
of PNC ECs which is consistent with the DDH potential was derived
in \cite{Liu:2003au}, Eqs. (17--24). The whole evaluation is straightforward,
however tedious, so we defer all the analytical expressions in Appendix
\ref{sec:appendix-PNC-ECs} and only quote the numerical results here.

\begin{figure}
\includegraphics[%
  scale=0.3,
  angle=270]{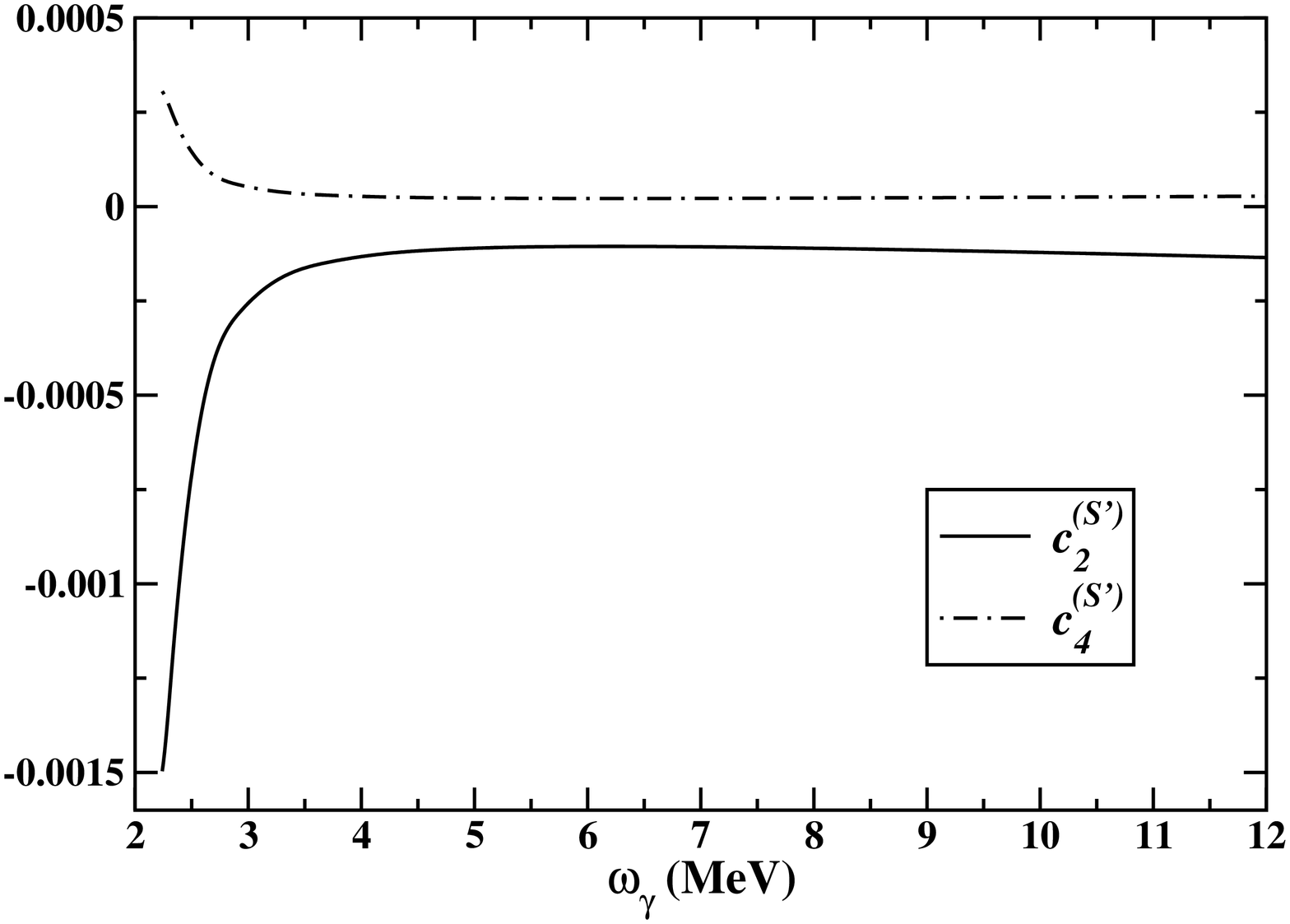}

\includegraphics[%
  scale=0.3,
  angle=270]{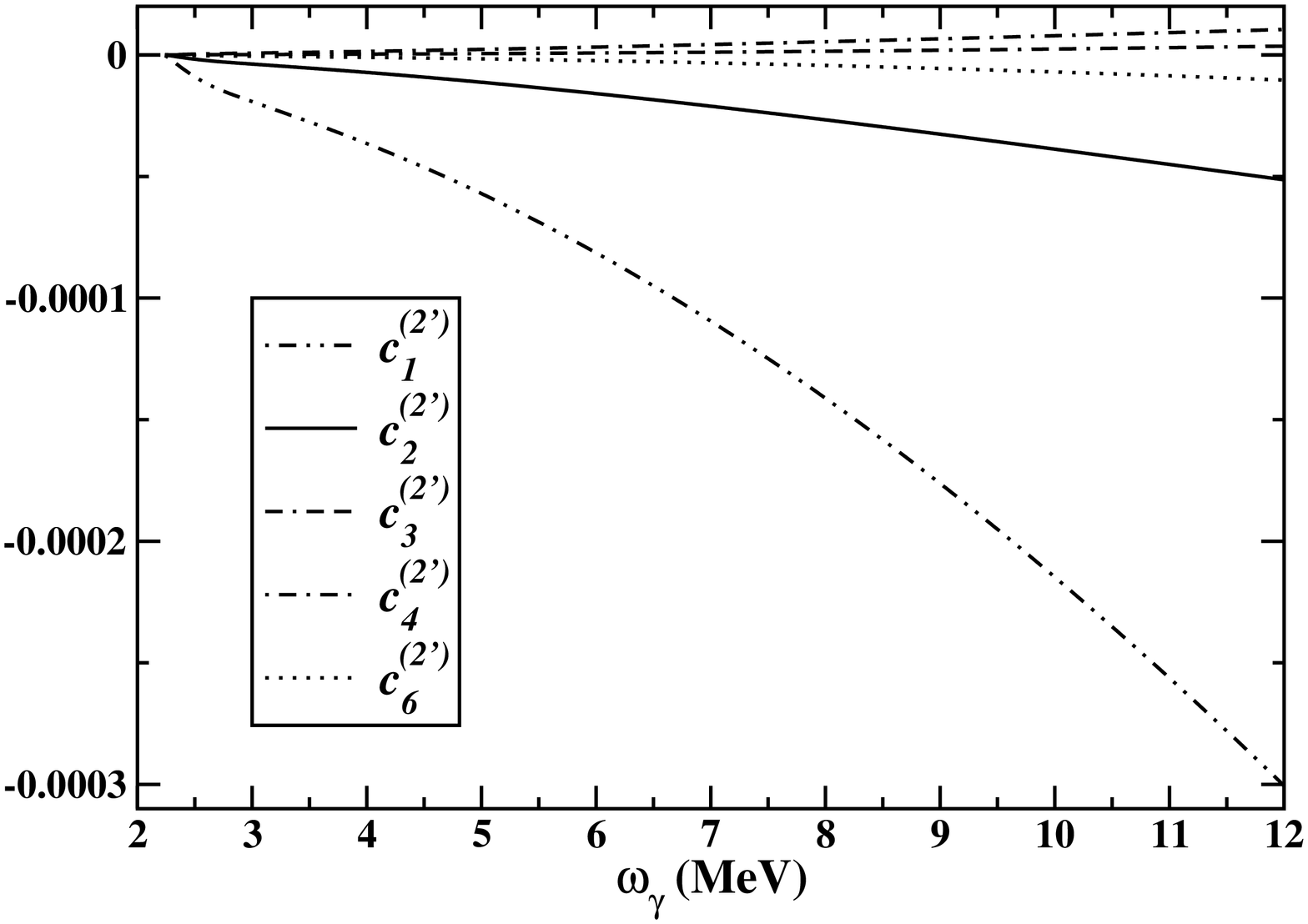}

\caption{The energy-dependences of PNC EC coefficients in the asymmetry parametrization:
the top panel shows $c_{2,4}^{(S')}$ in Eq. (\ref{eq:E1-PNC}) and
the bottom panel shows $c_{1,2,3,4,6}^{(2')}$ of Eq. (\ref{eq:M1-PNC}).
\label{fig:pncecs}}
\end{figure}

With the same parametrization as Eq. (\ref{eq:A exp}), the additional
contributions to the asymmetry by PNC ECs, via $E1_{5}$ and $M1_{5}$
respectively, are \begin{eqnarray}
A_{\gamma}(\tilde{F}_{E1_{5}}^{(S')}) & = & c_{2}^{(S')}\, h_{\rho}^{0}+c_{4}^{(S')}\, h_{\rho}^{2}\,,\label{eq:E1-PNC}\\
A_{\gamma}(\tilde{F}_{M1_{5}}^{(2')}) & = & c_{1}^{(2')}\, h_{\pi}^{1}+c_{2}^{(2')}\, h_{\rho}^{0}+c_{3}^{(2')}\, h_{\rho}^{1}\nonumber \\
 &  & +c_{4}^{(2')}\, h_{\rho}^{2}+c_{6}^{(2')}\, h_{\omega}^{1}\,.\label{eq:M1-PNC}\end{eqnarray}
 The detailed energy-dependence of each coefficient is shown in Fig.
\ref{fig:pncecs}. %

The dominance of $\tilde{F}_{E1_{5}}^{(S')}$ near the threshold and
$\tilde{F}_{M1_{5}}^{(2')}$ at higher energies could be readily observed
in these plots. We discuss their significances to the total asymmetry
in the following.

For the case where the photon energy is 0.01 MeV above the threshold,
only $c_{2}^{(S')}$ and $c_{4}^{(S')}$ are substantial. The former
coefficient is about 20\% of $c_{2}$, while the latter one is only
2\% of $c_{4}$. By using the DDH best values, these contributions
give an asymmetry about 1.4$\times10^{-9}$, which is a 6\% correction.
This is typically the order of magnitude one could expect from the
exchange effects.

As the energy gets larger, while the coefficients $c_{2}^{(S')}$
and $c_{4}^{(S')}$ keep stable, the coefficients associated with
$M1_{5}$ matrix elements grow linearly, roughly. The fastest growing
one is $c_{1}^{(2')}$ because the long-ranged pion-exchange dominates
the matrix elements. Comparatively, $c_{2}^{(2')}$ has a smaller
slope due to less overlap between the effective ranges pertinent to
the deuteron wave function and the $\rho$-exchange.

For the case where the photon energy is 10 MeV above the threshold,
$c_{1}^{(2')}$, $c_{2}^{(2')}$, and $c_{2}^{(S')}$ are substantial.
The first coefficient is about 50\% of $c_{1}$, and the latter two
combined is about 16\% of $c_{2}$. The extremely large correction
to $c_{1}$ can be simply explained. The cancellation which affects
the single-particle contribution (see Eq. (\ref{eq:cancellation}))
does not apply to the two-body one. By using the DDH best values,
all contributions due to PNC ECs give an asymmetry about 3.7$\times10^{-11}$,
which is a 5\% correction. The reason why large effects from individual
meson exchanges lead to an overall small correction is due to the
cancellation between pion and heavy-meson exchanges: the DDH best
values have opposite signs for the pion and heavy-meson couplings.
This conclusion however depends on the sign we assumed for the $g_{\rho\pi\gamma}$
coupling.

\section{Conclusion \label{sec:conclusion}}

The present work has been motivated by various aspects of the PNC
asymmetry $A_{\gamma}$ in the deuteron photodisintegration, especially
in the few-MeV photon-energy range. A first work addressing this energy
domain \cite{Oka:1983sp} showed that the process could provide information
on the PNC $\pi NN$ coupling constant, $h_{\pi}^{1}$, which allows
one to check results from other processes involving this coupling.
A later work \cite{Khriplovich:2000mb}, rather schematic, concluded
that this contribution could be largely suppressed. Between these
two extreme limits, the question arises of what this contribution
could be when a realistic description is made, including in particular
the tensor and spin-orbit components of the $NN$ interaction. At
the first sight, a sizable PNC $\pi$-exchange contribution could
arise if one assumes tensor-force effects of about 15\% for each partial
contribution and no cancellation.

The complete calculation shows that the $\pi$-exchange contribution
remains strongly suppressed after improving upon the schematic model.
Beyond making this observation, a genuine explanation should therefore
be found. When considering the asymmetry $A_{\gamma}$, an average
is made over the spins of initial and final states. Terms in the interference
effects of electric and magnetic transitions, whose spin dependence
averages to a non-zero value, are expected to produce a sizable contribution.
This discards the $\pi$-exchange contribution which involves a linear
dependence on the spin operator $\bm S$ and tensor-force effects
which involve the product of spin operators of order 1 and 2. The
argument applies to MECs too. A different conclusion would hold for
an observable implying a spin polarization of the initial or final
state. It thus appears some similarity between the relative role of
various contributions here and that one emphasized by Danilov for
the inverse process at thermal energies: the circular polarization
of photons $P_{\gamma}$ (equivalent to $A_{\gamma}$ here) is mainly
dependent on the PNC isoscalar and isotensor contributions while the
asymmetry of the photon emission with respect to the neutron polarization
depends on the $\pi$-exchange contribution.

As the $\pi$-exchange contribution to the asymmetry $A_{\gamma}$
turns out to have a minor role, we can concentrate on the vector-meson
ones. At the low energies considered here, it is expected that these
contributions depend on two combinations of parameters entering the
description of the PNC (and PC) $NN$ interaction. They are the zero-energy
neutron-proton scattering amplitudes in the $T=0$ and $T=1$ channels,
$\lambda_{t}$ and $\lambda_{s}$. In terms of these quantities introduced
by Danilov \cite{Danilov:1965} (see also later works by Missimer
\cite{Missimer:1976wb}, Desplanques and Missimer \cite{Desplanques:1978mt},
Holstein \cite{Holstein}) the discussion could be simpler. The asymmetry
is found to vary between \[
A_{\gamma}=0.70\,\mN\,\lambda_{t}-0.17\,\mN\,\lambda_{s}\;\;{\textrm{at \; threshold}}\]
 and \[
A_{\gamma}=-0.037\,\mN\,\lambda_{t}+0.022\,\mN\,\lambda_{s}\;\;{\textrm{at}}\;\omega_{\gamma}=12{\textrm{ MeV}},\]
 thus evidencing a change in sign which occurs around $\omega_{\gamma}=5.5\;{\textrm{MeV}}$
for both amplitudes. Depending on low-energy properties and, thus,
on the best known properties of the strong interaction, the place
where the cancellation of $A_{\gamma}$ occurs sounds to be well established.
It roughly agrees with what can be inferred from the analytic work
by Khriplovich and Korkin \cite{Khriplovich:2000mb}. Not much sensitivity
to PNC ECs is found. An experiment should therefore aim at a measurement
at energies significantly different, either below or above.

The goal for studying PNC effects is to get information on the hadronic
physics entering the PNC $NN$ interaction. This supposes that one
can disentangle the different contributions to each process. We notice
that the combination of parameters $\lambda_{t}$ and $\lambda_{s}$
appearing in the expression of $A_{\gamma}$ is orthogonal to that
one determining PNC effects in most other processes, especially in
medium and heavy nuclei. The study of the present process is therefore
quite useful. Another observation, which is not totally independent
of the previous one, concerns the isotensor contribution. This one
is especially favored in the present process while it is generally
suppressed in processes involving a roughly equal number of protons
and neutrons with either spin \cite{Desplanques:1998ak}. The present
process is therefore among the best ones to get information on the
isotensor $\rho NN$ coupling constant. We however stress that this
supposes the isoscalar parts could be constrained well by other processes.
In a meson-exchange model of the PNC interaction, these ones are represented
by the isoscalar $\rho NN$ and $\omega NN$ coupling constants. One
could add that the relative sign of these two contributions is the
same, in a large range of the photon energy ($\omega_{\gamma}\geq3\;$MeV),
as in many other processes. It however differs at small photon energies
where the asymmetry involves a combination of the various isoscalar
and isotensor couplings that is little constrained by other processes.
This explains that expectations of $A_{\gamma}$ up to $10^{-7}$
near threshold could be suggested in recent works on the basis of
a phenomenological analysis \cite{Desplanques:1998ak,Khriplovich:2000mb,Schi-int}.
Measuring this asymmetry could therefore be quite useful to determine
a poorly known component of PNC $NN$ interactions. On the theoretical
side, the present work should be completed by the contribution of
further parity-conserving exchange currents, but also by higher $1/m_{N}$-order
corrections from the single-particle current and, consistently, from
both PC and PNC exchange currents \cite{Friar:1983wd}. Though they
are not expected to change the main conclusions reached here, they
could be required to obtain from experiment a more accurate information
on PNC $NN$ forces.

\begin{acknowledgments}
C.-P.L. would like to thank R. Schiavilla, M. Fujiwara, and A.I. Titov
for useful disussions. C.H.H. gratefully acknowledges the hospitality
of the Laboratoire de Physique Subatomique et de Cosmologie, where
part of this work was performed. Work of C.H.H. is partially supported
by Korea Research Foundation (Grant No. KRF-2003-070-C00015).
\end{acknowledgments}
\appendix*

\section{Non-vanishing Matrix Elements of PNC ECs for Dominant Transitions
\label{sec:appendix-PNC-ECs}}

In this section, we summarize the analytical expressions of the non-zero
$\tilde{F}_{E1_{5}}^{(S')}$ and $\tilde{F}_{M1_{5}}^{(2')}$ for
the five dominant channels which lead to the numerical results in
Section \ref{sub:PNC-ECs}.\begin{widetext}

\subsection{$\tilde{F}_{E1_{5}}^{(S')}$}

As discussed in Section \ref{sec:formalism}, an exchange charge at
$O(1)$ should contribute to this form factor. According to Ref. \cite{Liu:2003au},
the $\rho$-exchange does generate one:\begin{equation}
\rho_{mesonic}^{\rho}(\bm x;\bm r_{1},\bm r_{2})=2\, e\, g_{\rho\ssst{NN}}\left(h_{\rho}^{0}-\frac{h_{\rho}^{2}}{2\sqrt{6}}\right)(\bm\tau_{1}\times\bm\tau_{2})^{z}(\bm\sigma_{1}-\bm\sigma_{2})\cdot\bm\nabla_{x}\Big(f_{\rho}(r_{x1})f_{\rho}(r_{x2})\Big)\,,\end{equation}
with $f_{\ssst{X}}(r)=\exp(-m_{\ssst{X}}\, r)/(4\,\pi\, r)$ and $r_{xi}=|\bm x-\bm r_{i}|$.
The $^{1}S_{0}$ state is the only open exit channel and it gives\begin{equation}
\langle E1_{5}^{(S')}\rangle=8\,\frac{g_{\rho\ssst{NN}}}{m_{\rho}}\left(h_{\rho}^{0}-\frac{h_{\rho}^{2}}{2\sqrt{6}}\right)\bra{^{1}S_{0}}r\, f_{\rho}(r)\ket{^{3}S_{1}}_{d}\,,\end{equation}
where $\bra{f}F(r)\ket{i}_{d}$ denotes the radial integral $\int dr\, U^{*}(f)\, F(r)\, U_{d}(i)$
and the subscript {}``$d$'' refers to the deuteron state.

\subsection{$\tilde{F}_{M1_{5}}^{(2')}$}

As the four allowed exit channels $^{3}P_{0}$, $^{3}P_{1}$ and $^{3}P_{2}$--$^{3}F_{2}$
are spin- and isospin-triplet, the non-vanishing PNC ECs, which satisfy
the spin and isospin selections rules, are \begin{eqnarray}
\bm j_{pair}^{\rho}(\bm x;\bm r_{1},\bm r_{2}) & = & \frac{e\, g_{\rho\ssst{NN}}}{4\mN}\, h_{\rho}^{1}\, f_{\rho}(r)(\tau_{1}^{z}-\tau_{2}^{z})(\bm\sigma_{1}+\bm\sigma_{2})\Big((1+\tau_{1}^{z})\delta^{(3)}(\bm x-\bm r_{1})\Big)+(1\leftrightarrow2)\,,\\
\bm j_{pair}^{\omega}(\bm x;\bm r_{1},\bm r_{2}) & = & \frac{-e\, g_{\omega\ssst{NN}}}{4\mN}\, h_{\omega}^{1}\, f_{\omega}(r)(\tau_{1}^{z}-\tau_{2}^{z})(\bm\sigma_{1}+\bm\sigma_{2})\Big((1+\tau_{1}^{z})\delta^{(3)}(\bm x-\bm r_{1})\Big)+(1\leftrightarrow2)\,,\\
\bm j_{mesonic}^{\rho}(\bm x;\bm r_{1},\bm r_{2}) & = & \frac{-e\, g_{\rho\ssst{NN}}}{\mN}\left(h_{\rho}^{0}-\frac{h_{\rho}^{2}}{2\sqrt{6}}\right)(\bm\tau_{1}\times\bm\tau_{2})^{z}\nabla_{x}^{a}\Big(i\left\{ \nabla_{1}^{a}\bm\sigma_{2}+\sigma_{1}^{a}\bm\nabla_{2},\, f_{\rho}(r_{x1})\, f_{\rho}(r_{x2})\right\} \nonumber \\
 &  & -\muV\left[(\bm\sigma_{1}\times\bm\nabla_{1})^{a}\bm\sigma_{2}+\sigma_{1}^{a}\bm\sigma_{2}\times\bm\nabla_{2},\, f_{\rho}(r_{x1})\, f_{\rho}(r_{x2})\right]\Big)+(1\leftrightarrow2)\,,\\
\bm j_{mesonic}^{\rho\pi}(\bm x;\bm r_{1},\bm r_{2}) & = & \frac{-e\, g_{\rho\ssst{NN}}\, g_{\rho\pi\gamma}}{\sqrt{2}\, m_{\rho}}\, h_{\pi}^{1}(\bm\tau_{1}\times\bm\tau_{2})^{z}(\bm\nabla_{1}\times\bm\nabla_{2})\Big(f_{\rho}(r_{x1})f_{\pi}(r_{x2})\Big)+(1\leftrightarrow2)\,.\label{eq:rhopiMesonic}\end{eqnarray}
 Note that one additional strong meson-nucleon coupling constant,
$g_{\rho\pi\gamma}$, appears in Eq. (\ref{eq:rhopiMesonic}). This
could be constrained by the $\rho\rightarrow\pi+\gamma$ data. For
the numerical calculation, we quote the number $g_{\rho\pi\gamma}=0.585$
as given in Ref. \cite{Truhlik:2000yx}. The matrix element $\langle M1_{5}^{(2')}\rangle$
can be written as a sum of the contributions from each EC as\begin{equation}
\langle M1_{5}^{(2')}\rangle=\frac{1}{\mN}\left(g_{\rho\ssst{NN}}\, h_{\rho}^{1}\, X_{1}+g_{\omega\ssst{NN}}\, h_{\omega}^{1}\, X_{2}+g_{\rho\ssst{NN}}\,\left(h_{\rho}^{0}-\frac{h_{\rho}^{2}}{2\sqrt{6}}\right)X_{3}\right)+\frac{1}{m_{\rho}}\,\grhoNN\, g_{\rho\pi\gamma}\, h_{\pi}^{1}\, X_{4}\,,\end{equation}
 and for each exit channel, the quantities $X_{1,2,3,4}$ are 

1. $^{3}P_{0}$\begin{eqnarray}
X_{1} & = & -\frac{2}{3}\left(\bra{^{3}P_{0}}r\, f_{\rho}(r)\ket{^{3}S_{1}}_{d}+\frac{1}{\sqrt{2}}\,\bra{^{3}P_{0}}r\, f_{\rho}(r)\ket{^{3}D_{1}}_{d}\right)\,,\\
X_{2} & = & \frac{2}{3}\left(\bra{^{3}P_{0}}r\, f_{\omega}(r)\ket{^{3}S_{1}}_{d}+\frac{1}{\sqrt{2}}\,\bra{^{3}P_{0}}r\, f_{\omega}(r)\ket{^{3}D_{1}}_{d}\right)\,,\\
X_{3} & = & -\frac{8}{3}\left((1+2\muV)\,\bra{^{3}P_{0}}r\, f_{\rho}(r)\ket{^{3}S_{1}}_{d}+\frac{1}{\sqrt{2}}\,(1-\muV)\,\bra{^{3}P_{0}}r\, f_{\rho}(r)\ket{^{3}D_{1}}_{d}\right.\nonumber \\
 &  & \left.\hspace{0.7cm}-\frac{2}{m_{\rho}}\,\bra{^{3}P_{0}}r\, f_{\rho}(r)\ket{^{3}S_{1}^{(+)}}_{d}-\frac{\sqrt{2}}{m_{\rho}}\,\bra{^{3}P_{0}}r\, f_{\rho}(r)\ket{^{3}D_{1}^{(-)}}_{d}\right),\\
X_{4} & = & \frac{4\sqrt{2}}{3\,(m_{\rho}^{2}-m_{\pi}^{2})}\left(\bra{^{3}P_{0}}f_{\pi\rho}^{\,'}(r)\ket{^{3}S_{1}}_{d}-\sqrt{2}\,\bra{^{3}P_{0}}f_{\pi\rho}^{\,'}(r)\ket{^{3}D_{1}}_{d}\right)\,;\end{eqnarray}

2. $^{3}P_{1}$

\begin{eqnarray}
X_{1} & = & \frac{1}{\sqrt{3}}\left(\bra{^{3}P_{1}}r\, f_{\rho}(r)\ket{^{3}S_{1}}_{d}-\sqrt{2}\,\bra{^{3}P_{1}}r\, f_{\rho}(r)\ket{^{3}D_{1}}_{d}\right)\,,\\
X_{2} & = & -\frac{1}{\sqrt{3}}\left(\bra{^{3}P_{1}}r\, f_{\omega}(r)\ket{^{3}S_{1}}_{d}-\sqrt{2}\,\bra{^{3}P_{1}}r\, f_{\omega}(r)\ket{^{3}D_{1}}_{d}\right)\,,\\
X_{3} & = & \frac{4}{\sqrt{3}}\bigg((1-\muV)\,\bra{^{3}P_{1}}r\, f_{\rho}(r)\ket{^{3}S_{1}}_{d}-\sqrt{2}\,(1-\muV)\,\bra{^{3}P_{1}}r\, f_{\rho}(r)\ket{^{3}D_{1}}_{d}\nonumber \\
 &  & \hspace{0.7cm}-\frac{2}{m_{\rho}}\,\bra{^{3}P_{1}}r\, f_{\rho}(r)\ket{^{3}S_{1}^{(+)}}_{d}+\frac{2\sqrt{2}}{m_{\rho}}\,\bra{^{3}P_{1}}r\, f_{\rho}(r)\ket{^{3}D_{1}^{(-)}}_{d}\bigg),\\
X_{4} & = & -\frac{4\sqrt{2}}{\sqrt{3}\,(m_{\rho}^{2}-m_{\pi}^{2})}\left(\bra{^{3}P_{1}}f_{\pi\rho}^{\,'}(r)\ket{^{3}S_{1}}_{d}+\frac{1}{\sqrt{2}}\,\bra{^{3}P_{1}}f_{\pi\rho}^{\,'}(r)\ket{^{3}D_{1}}_{d}\right)\,;\end{eqnarray}

3. $^{3}P_{2}$--$^{3}F_{2}$

\begin{eqnarray}
X & = & \frac{\sqrt{5}}{3}\left(\bra{^{3}P_{2}}r\, f_{\rho}(r)\ket{^{3}S_{1}}_{d}-\frac{2\sqrt{2}}{5}\,\bra{^{3}P_{2}}r\, f_{\rho}(r)\ket{^{3}D_{1}}_{d}-\frac{3\sqrt{3}}{5}\,\bra{^{3}F_{2}}r\, f_{\rho}(r)\ket{^{3}D_{1}}_{d}\right)\,,\\
X_{2} & = & -\frac{\sqrt{5}}{3}\left(\bra{^{3}P_{2}}r\, f_{\omega}(r)\ket{^{3}S_{1}}_{d}\,-\frac{2\sqrt{2}}{5}\bra{^{3}P_{2}}r\, f_{\omega}(r)\ket{^{3}D_{1}}_{d}-\frac{3\sqrt{3}}{5}\,\bra{^{3}F_{2}}r\, f_{\omega}(r)\ket{^{3}D_{1}}_{d}\right)\,,\\
X_{3} & = & \frac{4\sqrt{5}}{3}\left((1-\muV)\,\bra{^{3}P_{2}}r\, f_{\rho}(r)\ket{^{3}S_{1}}_{d}-\frac{2\sqrt{2}}{5}\,(1-4\muV)\,\bra{^{3}P_{2}}r\, f_{\rho}(r)\ket{^{3}D_{1}}_{d}\right.\nonumber \\
 &  & \hspace{1.0cm}-\frac{2}{m_{\rho}}\,\bra{^{3}P_{2}}r\, f_{\rho}(r)\ket{^{3}S_{1}^{(+)}}_{d}+\frac{4\sqrt{2}}{5\, m_{\rho}}\,\bra{^{3}P_{2}}r\, f_{\rho}(r)\ket{^{3}D_{1}^{(-)}}_{d}\nonumber \\
 &  & \left.\hspace{1.0cm}-\frac{3\sqrt{3}}{5}\,(1+\muV)\,\bra{^{3}F_{2}}r\, f_{\rho}(r)\ket{^{3}D_{1}}_{d}+\frac{6\sqrt{3}}{5\, m_{\rho}}\,\bra{^{3}F_{2}}r\, f_{\rho}(r)\ket{^{3}D_{1}^{(+)}}_{d}\right),\\
X_{4} & = & \frac{4\sqrt{10}}{3\,(m_{\rho}^{2}-m_{\pi}^{2})}\left(\bra{^{3}P_{2}}f_{\pi\rho}^{\,'}(r)\ket{^{3}S_{1}}_{d}-\frac{1}{5\sqrt{2}}\,\bra{^{3}P_{2}}f_{\pi\rho}^{\,'}(r)\ket{^{3}D_{1}}_{d}+\frac{3\sqrt{3}}{5}\,\bra{^{3}F_{2}}f_{\pi\rho}^{\,'}(r)\ket{^{3}D_{1}}_{d}\right)\,,\end{eqnarray}
where $\ket{^{2S+1}L_{J}^{(+)}}\equiv\left(\frac{d}{dr}-\frac{L+1}{r}\right)\ket{^{2S+1}L_{J}}$,
$\ket{^{2S+1}L_{J}^{(-)}}\equiv\left(\frac{d}{dr}+\frac{L}{r}\right)\ket{^{2S+1}L_{J}}$,
and $f_{\pi\rho}^{\,'}(r)\equiv\frac{d}{dr}\left(f_{\pi}(r)-f_{\rho}(r)\right)$.

\bibliographystyle{apsrev}
\bibliography{QEPV,MEC,AM,npPV}

\end{widetext}

\end{document}

%% file: my_macro.tex
\newcommand{\ssst}[1]{\scriptscriptstyle{#1}}

\newcommand{\bra}[1]{\langle #1 \vert}
\newcommand{\ket}[1]{\vert #1 \rangle}

\newcommand{\mN}{m_{\ssst{N}}}

\newcommand{\grhoNN}{g_{\rho\ssst{NN}}}

\newcommand{\muS}{\mu_{\ssst{S}}}
\newcommand{\muV}{\mu_{\ssst{V}}}

%% file: gdnp_2c.bbl
\begin{thebibliography}{31}
\expandafter\ifx\csname natexlab\endcsname\relax\def\natexlab#1{#1}\fi
\expandafter\ifx\csname bibnamefont\endcsname\relax
  \def\bibnamefont#1{#1}\fi
\expandafter\ifx\csname bibfnamefont\endcsname\relax
  \def\bibfnamefont#1{#1}\fi
\expandafter\ifx\csname citenamefont\endcsname\relax
  \def\citenamefont#1{#1}\fi
\expandafter\ifx\csname url\endcsname\relax
  \def\url#1{\texttt{#1}}\fi
\expandafter\ifx\csname urlprefix\endcsname\relax\def\urlprefix{URL }\fi
\providecommand{\bibinfo}[2]{#2}
\providecommand{\eprint}[2][]{\url{#2}}

\bibitem[{\citenamefont{Danilov}(1965)}]{Danilov:1965}
\bibinfo{author}{\bibfnamefont{G.~S.} \bibnamefont{Danilov}},
  \bibinfo{journal}{Phys. Lett.} \textbf{\bibinfo{volume}{18}},
  \bibinfo{pages}{40} (\bibinfo{year}{1965}).

\bibitem[{\citenamefont{Lobashov~\etal}(1972)}]{Lobashov:1972}
\bibinfo{author}{\bibfnamefont{V.~M.} \bibnamefont{Lobashov~\etal}},
  \bibinfo{journal}{Nucl. Phys. A} \textbf{\bibinfo{volume}{197}},
  \bibinfo{pages}{241} (\bibinfo{year}{1972}).

\bibitem[{\citenamefont{Lassey and McKellar}(1975)}]{Lassey:1975}
\bibinfo{author}{\bibfnamefont{K.~R.} \bibnamefont{Lassey}} \bibnamefont{and}
  \bibinfo{author}{\bibfnamefont{B.~H.~J.} \bibnamefont{McKellar}},
  \bibinfo{journal}{Phys. Rev. C} \textbf{\bibinfo{volume}{11}},
  \bibinfo{pages}{349} (\bibinfo{year}{1975}).

\bibitem[{\citenamefont{Desplanques}(1975)}]{Desplanques:1975}
\bibinfo{author}{\bibfnamefont{B.}~\bibnamefont{Desplanques}},
  \bibinfo{journal}{Nucl. Phys. A} \textbf{\bibinfo{volume}{242}},
  \bibinfo{pages}{423} (\bibinfo{year}{1975}).

\bibitem[{\citenamefont{Craver et~al.}(1976)\citenamefont{Craver, Fischbach,
  Kim, and Tubis}}]{Craver:1976am}
\bibinfo{author}{\bibfnamefont{B.~A.} \bibnamefont{Craver}},
  \bibinfo{author}{\bibfnamefont{E.}~\bibnamefont{Fischbach}},
  \bibinfo{author}{\bibfnamefont{Y.~E.} \bibnamefont{Kim}}, \bibnamefont{and}
  \bibinfo{author}{\bibfnamefont{A.}~\bibnamefont{Tubis}},
  \bibinfo{journal}{Phys. Rev. D} \textbf{\bibinfo{volume}{13}},
  \bibinfo{pages}{1376} (\bibinfo{year}{1976}).

\bibitem[{\citenamefont{Knyazkov~\etal}(1983)}]{Knyazkov:1983ke}
\bibinfo{author}{\bibfnamefont{V.~A.} \bibnamefont{Knyazkov~\etal}},
  \bibinfo{journal}{JETP Lett.} \textbf{\bibinfo{volume}{38}},
  \bibinfo{pages}{163} (\bibinfo{year}{1983}).

\bibitem[{\citenamefont{Knyazkov~\etal}(1984)}]{Knyazkov:1984}
\bibinfo{author}{\bibfnamefont{V.~A.} \bibnamefont{Knyazkov~\etal}},
  \bibinfo{journal}{Nucl. Phys. A} \textbf{\bibinfo{volume}{417}},
  \bibinfo{pages}{209} (\bibinfo{year}{1984}).

\bibitem[{\citenamefont{Berdoz~\etal}(2001)}]{Berdoz:2001nu}
\bibinfo{author}{\bibfnamefont{A.~R.} \bibnamefont{Berdoz~\etal}},
  \bibinfo{journal}{Phys. Rev. Lett.} \textbf{\bibinfo{volume}{87}},
  \bibinfo{pages}{272301} (\bibinfo{year}{2001}).

\bibitem[{\citenamefont{Snow~\etal}(2000)}]{Snow:2000az}
\bibinfo{author}{\bibfnamefont{W.~M.} \bibnamefont{Snow~\etal}},
  \bibinfo{journal}{Nucl. Instrum. Meth. A} \textbf{\bibinfo{volume}{440}},
  \bibinfo{pages}{729} (\bibinfo{year}{2000}).

\bibitem[{\citenamefont{Hasty~\etal}(2000)}]{SAMPLE00b}
\bibinfo{author}{\bibfnamefont{R.}~\bibnamefont{Hasty~\etal}},
  \bibinfo{journal}{Science} \textbf{\bibinfo{volume}{290}},
  \bibinfo{pages}{2117} (\bibinfo{year}{2000}).

\bibitem[{\citenamefont{Ito~\etal}(2003)}]{Ito:2003mr}
\bibinfo{author}{\bibfnamefont{T.~M.} \bibnamefont{Ito~\etal}}
  (\bibinfo{year}{2003}), \eprint{nucl-ex/0310001}.

\bibitem[{\citenamefont{Lee}(1978)}]{Lee:1978kh}
\bibinfo{author}{\bibfnamefont{H.~C.} \bibnamefont{Lee}},
  \bibinfo{journal}{Phys. Rev. Lett.} \textbf{\bibinfo{volume}{41}},
  \bibinfo{pages}{843} (\bibinfo{year}{1978}).

\bibitem[{\citenamefont{Oka}(1983)}]{Oka:1983sp}
\bibinfo{author}{\bibfnamefont{T.}~\bibnamefont{Oka}}, \bibinfo{journal}{Phys.
  Rev. D} \textbf{\bibinfo{volume}{27}}, \bibinfo{pages}{523}
  (\bibinfo{year}{1983}).

\bibitem[{\citenamefont{Khriplovich and Korkin}(2001)}]{Khriplovich:2000mb}
\bibinfo{author}{\bibfnamefont{I.~B.} \bibnamefont{Khriplovich}}
  \bibnamefont{and} \bibinfo{author}{\bibfnamefont{R.~V.}
  \bibnamefont{Korkin}}, \bibinfo{journal}{Nucl. Phys. A}
  \textbf{\bibinfo{volume}{690}}, \bibinfo{pages}{610} (\bibinfo{year}{2001}).

\bibitem[{\citenamefont{Earle et~al.}(1981)\citenamefont{Earle, McDonald, and
  Knowles}}]{Earle:1981}
\bibinfo{author}{\bibfnamefont{E.~D.} \bibnamefont{Earle}},
  \bibinfo{author}{\bibfnamefont{A.~B.} \bibnamefont{McDonald}},
  \bibnamefont{and} \bibinfo{author}{\bibfnamefont{J.~W.}
  \bibnamefont{Knowles}}, in \emph{\bibinfo{booktitle}{AIP Conf. Proc.}}
  (\bibinfo{year}{1981}), vol.~\bibinfo{volume}{69}, p. \bibinfo{pages}{1436}.

\bibitem[{\citenamefont{Earle~\etal}(1988)}]{Earle:1988fc}
\bibinfo{author}{\bibfnamefont{E.~D.} \bibnamefont{Earle~\etal}},
  \bibinfo{journal}{Can. J. Phys.} \textbf{\bibinfo{volume}{66}},
  \bibinfo{pages}{534} (\bibinfo{year}{1988}).

\bibitem[{\citenamefont{Wojtsekhowski and van Oers}()}]{jlab-lett00}
\bibinfo{author}{\bibfnamefont{B.}~\bibnamefont{Wojtsekhowski}}
  \bibnamefont{and} \bibinfo{author}{\bibfnamefont{W.~T.~H.} \bibnamefont{van
  Oers}}, \bibinfo{note}{{JLAB} letter-of-intent 00-002}.

\bibitem[{\citenamefont{Siegert}(1937)}]{Siegert:1937yt}
\bibinfo{author}{\bibfnamefont{A.~J.~F.} \bibnamefont{Siegert}},
  \bibinfo{journal}{Phys. Rev.} \textbf{\bibinfo{volume}{52}},
  \bibinfo{pages}{787} (\bibinfo{year}{1937}).

\bibitem[{\citenamefont{Desplanques et~al.}(1980)\citenamefont{Desplanques,
  Donoghue, and Holstein}}]{Desplanques:1980hn}
\bibinfo{author}{\bibfnamefont{B.}~\bibnamefont{Desplanques}},
  \bibinfo{author}{\bibfnamefont{J.~F.} \bibnamefont{Donoghue}},
  \bibnamefont{and} \bibinfo{author}{\bibfnamefont{B.~R.}
  \bibnamefont{Holstein}}, \bibinfo{journal}{Ann. Phys.}
  \textbf{\bibinfo{volume}{124}}, \bibinfo{pages}{449} (\bibinfo{year}{1980}).

\bibitem[{\citenamefont{Musolf~\etal}(1994)}]{Musolf:1994tb}
\bibinfo{author}{\bibfnamefont{M.~J.} \bibnamefont{Musolf~\etal}},
  \bibinfo{journal}{Phys. Rep.} \textbf{\bibinfo{volume}{239}},
  \bibinfo{pages}{1} (\bibinfo{year}{1994}).

\bibitem[{\citenamefont{Liu et~al.}(2003{\natexlab{a}})\citenamefont{Liu,
  Pr{\'e}zeau, and Ramsey-Musolf}}]{Liu:2002bq}
\bibinfo{author}{\bibfnamefont{C.-P.} \bibnamefont{Liu}},
  \bibinfo{author}{\bibfnamefont{G.}~\bibnamefont{Pr{\'e}zeau}},
  \bibnamefont{and} \bibinfo{author}{\bibfnamefont{M.~J.}
  \bibnamefont{Ramsey-Musolf}}, \bibinfo{journal}{Phys. Rev. C}
  \textbf{\bibinfo{volume}{67}}, \bibinfo{pages}{035501}
  (\bibinfo{year}{2003}{\natexlab{a}}).

\bibitem[{\citenamefont{Liu et~al.}(2003{\natexlab{b}})\citenamefont{Liu, Hyun,
  and Desplanques}}]{Liu:2003au}
\bibinfo{author}{\bibfnamefont{C.-P.} \bibnamefont{Liu}},
  \bibinfo{author}{\bibfnamefont{C.~H.} \bibnamefont{Hyun}}, \bibnamefont{and}
  \bibinfo{author}{\bibfnamefont{B.}~\bibnamefont{Desplanques}},
  \bibinfo{journal}{Phys. Rev. C} \textbf{\bibinfo{volume}{68}},
  \bibinfo{pages}{045501} (\bibinfo{year}{2003}{\natexlab{b}}).

\bibitem[{\citenamefont{Wiringa et~al.}(1995)\citenamefont{Wiringa, Stoks, and
  Schiavilla}}]{Wiringa:1995wb}
\bibinfo{author}{\bibfnamefont{R.~B.} \bibnamefont{Wiringa}},
  \bibinfo{author}{\bibfnamefont{V.~G.~J.} \bibnamefont{Stoks}},
  \bibnamefont{and}
  \bibinfo{author}{\bibfnamefont{R.}~\bibnamefont{Schiavilla}},
  \bibinfo{journal}{Phys. Rev. C} \textbf{\bibinfo{volume}{51}},
  \bibinfo{pages}{38} (\bibinfo{year}{1995}).

\bibitem[{\citenamefont{Arenh{\"o}vel and Sanzone}(1991)}]{Arenhovel:1991}
\bibinfo{author}{\bibfnamefont{H.}~\bibnamefont{Arenh{\"o}vel}}
  \bibnamefont{and} \bibinfo{author}{\bibfnamefont{H.}~\bibnamefont{Sanzone}},
  \bibinfo{journal}{Few Body Syst. Suppl.} \textbf{\bibinfo{volume}{3}},
  \bibinfo{pages}{1} (\bibinfo{year}{1991}).

\bibitem[{\citenamefont{Missimer}(1976)}]{Missimer:1976wb}
\bibinfo{author}{\bibfnamefont{J.}~\bibnamefont{Missimer}},
  \bibinfo{journal}{Phys. Rev. C} \textbf{\bibinfo{volume}{14}},
  \bibinfo{pages}{347} (\bibinfo{year}{1976}).

\bibitem[{\citenamefont{Desplanques and Missimer}(1978)}]{Desplanques:1978mt}
\bibinfo{author}{\bibfnamefont{B.}~\bibnamefont{Desplanques}} \bibnamefont{and}
  \bibinfo{author}{\bibfnamefont{J.}~\bibnamefont{Missimer}},
  \bibinfo{journal}{Nucl. Phys. A} \textbf{\bibinfo{volume}{300}},
  \bibinfo{pages}{286} (\bibinfo{year}{1978}).

\bibitem[{\citenamefont{Holstein}()}]{Holstein}
\bibinfo{author}{\bibfnamefont{B.~R.} \bibnamefont{Holstein}},
  \urlprefix\url{http://mocha.phys.washington.edu/~int_talk/WorkShops/
  int_02_3/People/Holstein_B/}.

\bibitem[{\citenamefont{Desplanques}(1998)}]{Desplanques:1998ak}
\bibinfo{author}{\bibfnamefont{B.}~\bibnamefont{Desplanques}},
  \bibinfo{journal}{Phys. Rep.} \textbf{\bibinfo{volume}{297}},
  \bibinfo{pages}{1} (\bibinfo{year}{1998}).

\bibitem[{\citenamefont{Schiavilla}()}]{Schi-int}
\bibinfo{author}{\bibfnamefont{R.}~\bibnamefont{Schiavilla}},
  \urlprefix\url{http://mocha.phys.washington.edu/~int_talk/WorkShops/
  int_03_3/People/Schiavilla_R/}.

\bibitem[{\citenamefont{Friar and McKellar}(1983)}]{Friar:1983wd}
\bibinfo{author}{\bibfnamefont{J.~L.} \bibnamefont{Friar}} \bibnamefont{and}
  \bibinfo{author}{\bibfnamefont{B.~H.~J.} \bibnamefont{McKellar}},
  \bibinfo{journal}{Phys. Lett. B} \textbf{\bibinfo{volume}{123}},
  \bibinfo{pages}{284} (\bibinfo{year}{1983}).

\bibitem[{\citenamefont{Truhlik et~al.}(2001)\citenamefont{Truhlik, Smejkal,
  and Khanna}}]{Truhlik:2000yx}
\bibinfo{author}{\bibfnamefont{E.}~\bibnamefont{Truhlik}},
  \bibinfo{author}{\bibfnamefont{J.}~\bibnamefont{Smejkal}}, \bibnamefont{and}
  \bibinfo{author}{\bibfnamefont{F.~C.} \bibnamefont{Khanna}},
  \bibinfo{journal}{Nucl. Phys. A} \textbf{\bibinfo{volume}{689}},
  \bibinfo{pages}{741} (\bibinfo{year}{2001}).

\end{thebibliography}
